\documentclass[11pt]{article}
\usepackage{palatino}
\usepackage{a4p}
\usepackage{pennames}
\usepackage[dvips,xdvi]{graphicx}

\def\etal{\mbox{{\it et al.}}}

\newcommand{\ra}{\ensuremath{\rightarrow}}
\begin{document}
\begin{titlepage}
\rightline{OPAL Conference Report CR465}
\rightline{19 March 2001}
\vspace{3cm}
\begin{center}
{\Large\bf Search for the Standard Model Higgs Boson in OPAL}\\[2cm]

{\large\bf I.Nakamura}\\
{\it International Center for Elementaly Particle Physics,
  University of Tokyo, Tokyo 113-0033, Japan}
  
\end{center}

\vspace{3cm}

\begin{abstract}
  A search for the Standard Model Higgs boson with the OPAL detector
  at LEP was reported.
  The analysis was based on the full data sample
  collected at $\sqrt{s}\approx$~192--209~GeV in 1999 and 2000,
  corresponding to an integrated luminosity
  of approximately 426 pb$^{-1}$.
  The data shows an excess in a range of 95--120 GeV\@.
  The data slightly
  favour the hypothesis that a signal is present, but also that the
  data are consistent with the background hypothesis.
  A lower bound of 109.7~GeV is obtained on the
  Higgs boson mass at the 95\% confidence level.
\end{abstract}

\vspace{2cm}
\begin{center}
  Talk presented in \\
  XXXVI th Rencontres de Moriond
  Electroweak Interactions and Unified Theories, 10-17 March 2001.
\end{center}

\end{titlepage}

\section{Introcuction}
In year 2000, a number of improvements were made to the LEP collider.
It increased Higgs discovery potential up to approximately 115 GeV
if LEP combination were made.
In LEPC held in November 2000, an excess of candidates, which is
compatible with the Standard Model(SM) Higgs boson of 115 GeV\@,
was reported\,\cite{higgs_piktalk}.
The results of a search for the SM Higgs boson with
the OPAL detector at LEP was reported in this talk based on the
results described in published letter\,\cite{smletter}.

\section{The Analyses}\label{sect:analyses}
The data used in the analyses correspond to integrated luminosities 
of approximately 216 pb$^{-1}$ at 192--202 GeV, 80 pb$^{-1}$
at 203--206 GeV, and 130 pb$^{-1}$ at centre-of-mass energies ($\sqrt{s}$)
higher than 206 GeV.  
During 2000 (1999), data were taken at
$\sqrt{s}=200$-209 GeV (192-202 GeV) with a luminosity-weighted mean
$\sqrt{s}$ of 206.1 (197.6) GeV\@. 

The analyses follows our previous publication\cite{sm189}.
The search was performed in the following final states,
\PHz\PZz\ra{}b\=bq\=q
(four-jet), \PHz\PZz\ra{}b\=b\Pgn\Pagn\ (missing energy),
\PHz\PZz\ra{}b\=b\Pgtp\Pgtm\ and \Pgtp\Pgtm{}q\=q (tau),
\PHz\PZz\ra{}b\=b\Pep\Pem{} and \PHz\PZz\ra{}b\=b\Pgmp\Pgmm{}
(electron and muon).

In all channels, preselections followed by likelihood selections
were applied to select the Higgs signal. The likelihood consists
of b-tagging information, event shape variables, lepton
identification, (in)compatibilities to the signal(background) mass
hypothesis and so on. The likelihood selection were optimised using a
mixture of various masses of signal Monte Carlo samples. Hence, the
selection was not optimized to a particular mass.

\begin{figure}[htbp]
  \begin{center}
    \includegraphics[width=.55\textwidth]{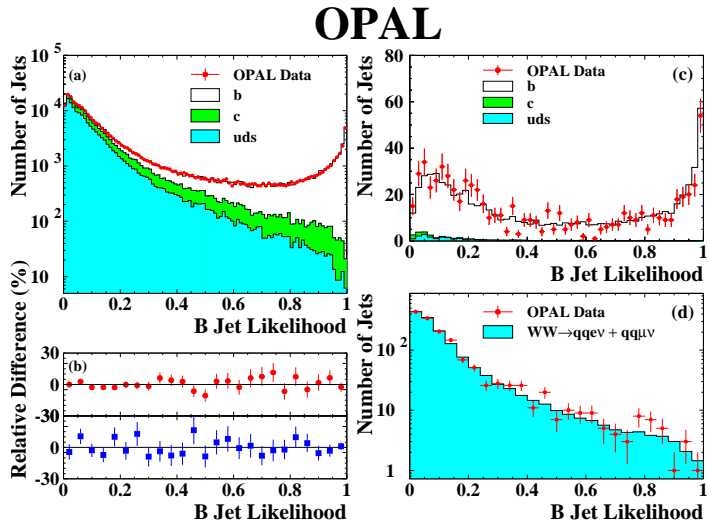}
    \caption{\sl
      The b-tagging performance and modeling for (a--b) calibration data 
      taken at \PZz\ pole, and (c--d) at 200--209 GeV in 2000.
      (a) The distribution of the b-tagging variable $\mathcal{B}$ for jets
      (b) The bin-by-bin difference between data and Monte Carlo simulation
      for b-enriched(circles) and b-suppressed(squares) samples.
      (c) The b-tagging output for opposite
      b-tagged jets in a sample of q\=q\Pgg\ events, and (d) for jets
      in a sample of \PWp\PWm$\to\ell\nu$q\=q ($\ell=$e or \Pgn) events.}
    \label{fig:btag}
  \end{center}
\end{figure}

The b-tagging variable $\mathcal{B}$ is a likelihood combination of
three independent b-tagging information, lifetime sensitive part,
jet kinematics and lepton $p_t$ information. Lifetime sensitive part
is an artificial neural network consists of four input variables
including both displaced vertex and track impact parameter information.
The performance of b-tagging was checked using the calibration data
taken at \PZz\ peak in 2000, as well as higher energy data as shown
in Figure\,\ref{fig:btag}.

\begin{figure}[htbp]
  \begin{center}
    \includegraphics[width=.45\textwidth]{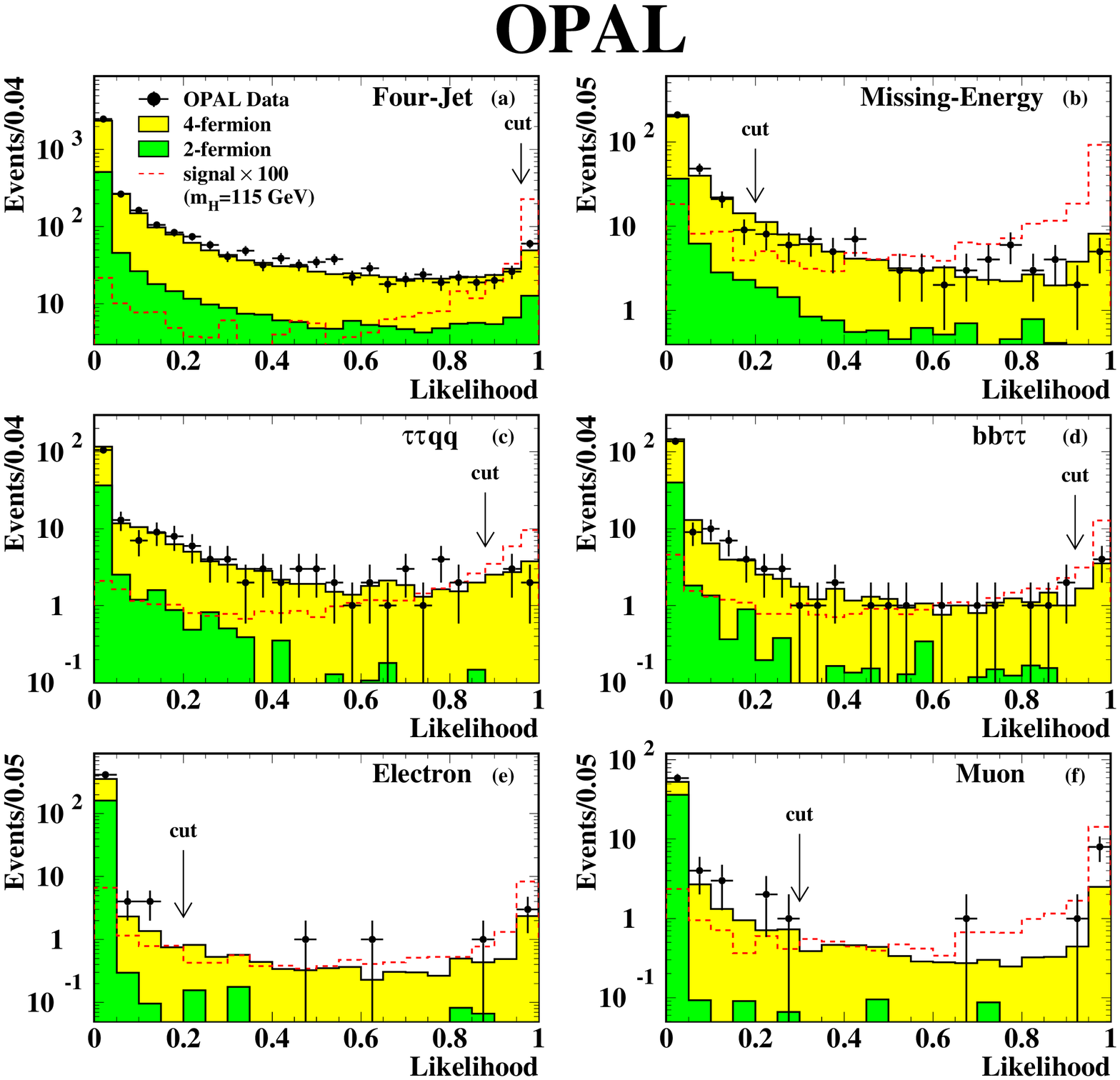}
    \includegraphics[width=.45\textwidth]{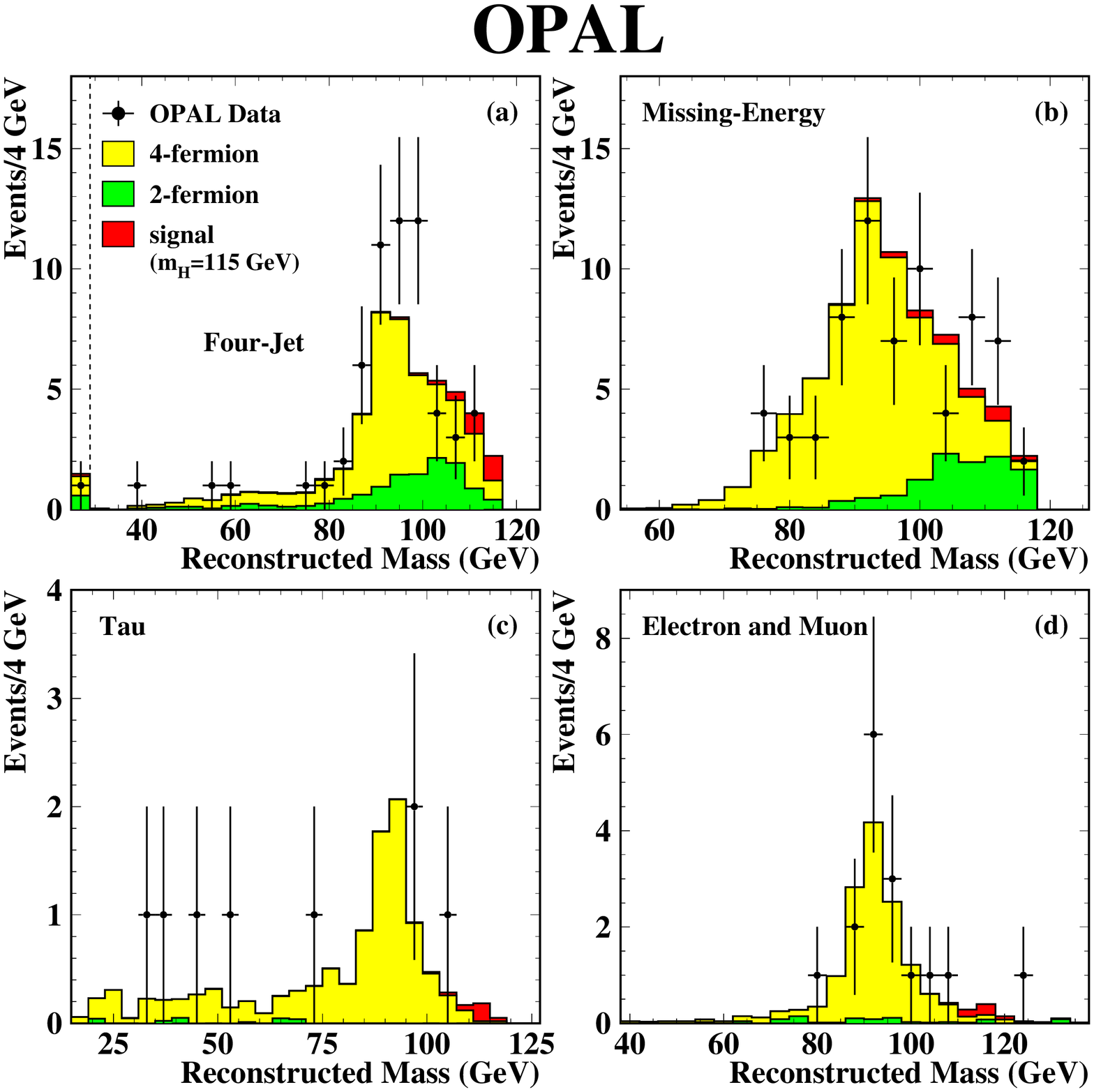}
    \caption{\sl
      Left:
      The likelihood distributions for each channels
      OPAL data are shown with points, 
      backgrounds with the shaded histograms, 
      and the expectation from a signal with $m_\PHz=$115 GeV with the
      dashed histograms 
      (scaled up by a factor of 100 for visibility.
      Right:
      The reconstructed mass distribution for the selected events}
    \label{fig:likelihood}
  \end{center}
\end{figure}

The number of selected events in all search channels 
was 156 with 146.1$\pm$11.9 expected from the SM background
processes.
Figure\,\ref{fig:likelihood} shows the likelihood distributions of each
channels after preselection and the reconstructed mass distributions
after likelihood selection.

After the event selections search channels were combined using
the likelihood ratio method\,\cite{likelihoodratio}
or the method described in previous paper\,\cite{sm189}.
Three different final
discriminants, ``test statistic'', were used in different channels
to reflect the different background conditions.
The four-jet channel uses a likelihood consists of four variables
including b-tagging, reconstructed mass and two event shape parameters.
The missing energy, electron and muon channel uses likelihood
calculated from mass and selection likelihood output. The tau
channel uses simply the mass distribution as a test statistic.

\section{Results}\label{sect:results}

\begin{figure}[htbp]
  \begin{center}
    \includegraphics[width=.40\textwidth]{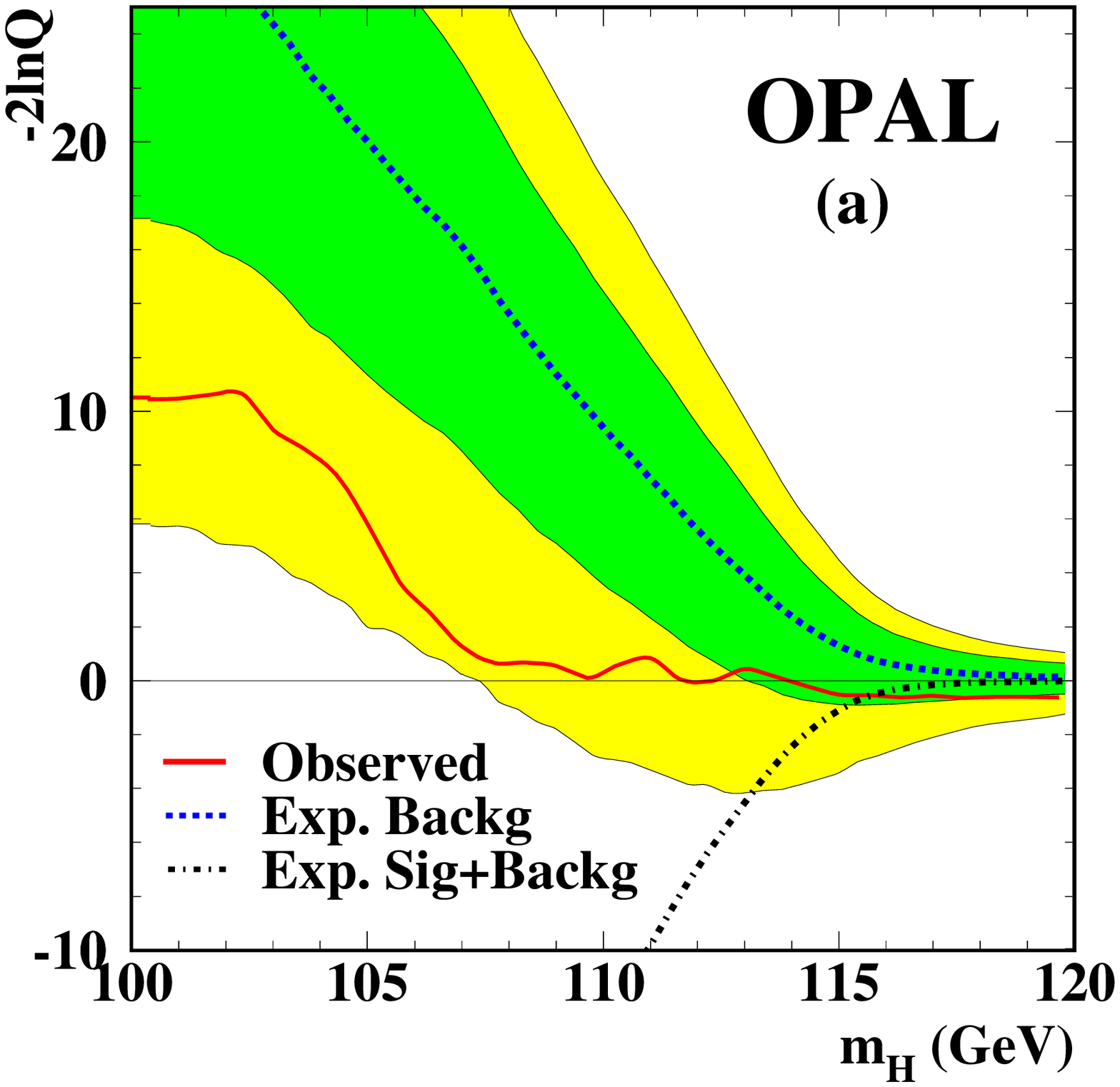}
    \includegraphics[width=.415\textwidth]{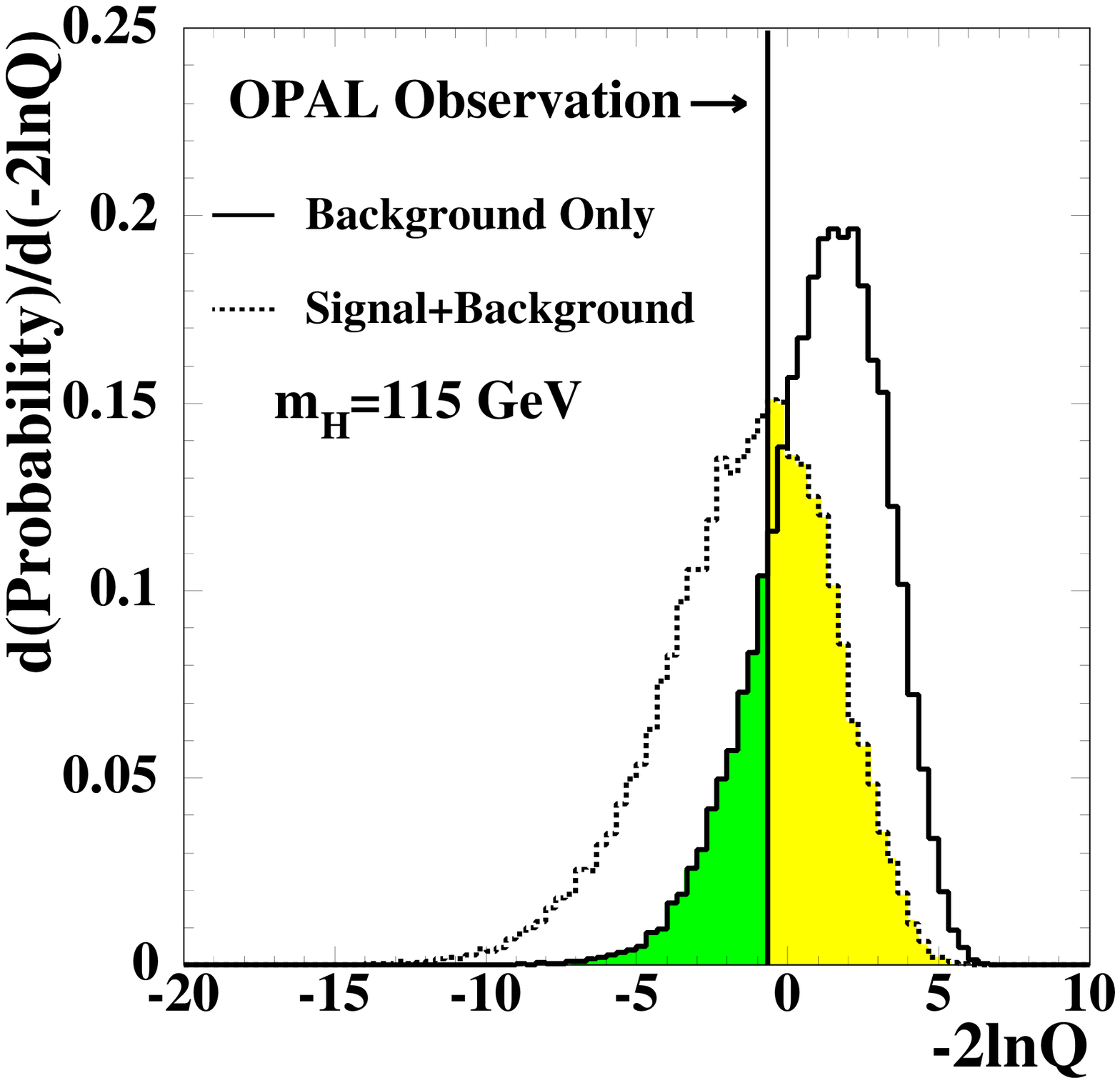}
    \caption{\sl
      (a) The log-likelihood ratio $-2\ln{\mathcal{Q}}$ as a function
      of the test mass $m_\PHz$.  
      The observation for the data 
      is shown with a solid line. 
      The dashed line indicates the median background expectation and
      the dark (light) shaded band shows the 68\% (95\%) probability
      intervals.
      The median expectation in the presence of a signal
      is shown with a dot-dashed line.
      (b) The expected $-2\ln{\mathcal{Q}}$ distribution for
      background only (solid line) and with a 115 GeV Higgs
      boson (dashed histogram).  The observation
      in the data is shown with a vertical solid line.}
    \label{fig:likelihoodratio}
  \end{center}
\end{figure}

Figure\,\ref{fig:likelihoodratio}(a) shows the log-likelihood ratio
as a function of test mass. The two bands in the figure are 68\% and
95\% contours centred on the median
expectation. Figure\,\ref{fig:likelihoodratio}(b) shows the projection
of (a) at test mass of 115 GeV\@, corresponding to the probability density
of $-2\ln{\mathcal{Q}}$ for signal+background and background only
hypotheses. The background confidence level (1$-$CL$_\mathrm{b}$) and
signal+background confidence level (CL$_\mathrm{s+b}$) are 0.2 and 0.4
shown in dark and light shaded area in the figure, respectively.

\begin{figure}[htbp]
  \begin{center}
    \includegraphics[width=.42\textwidth]{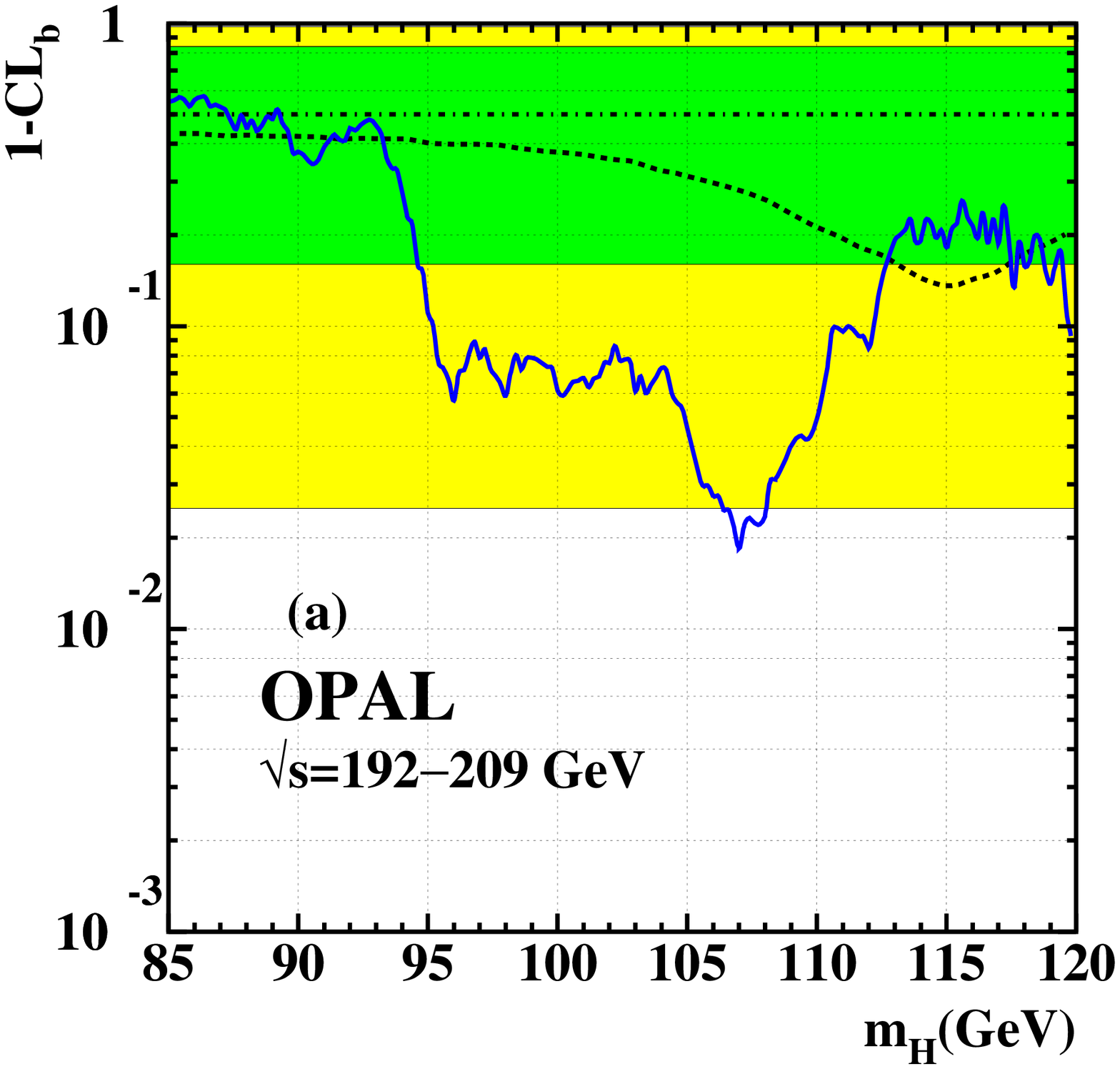}
    \includegraphics[width=.42\textwidth]{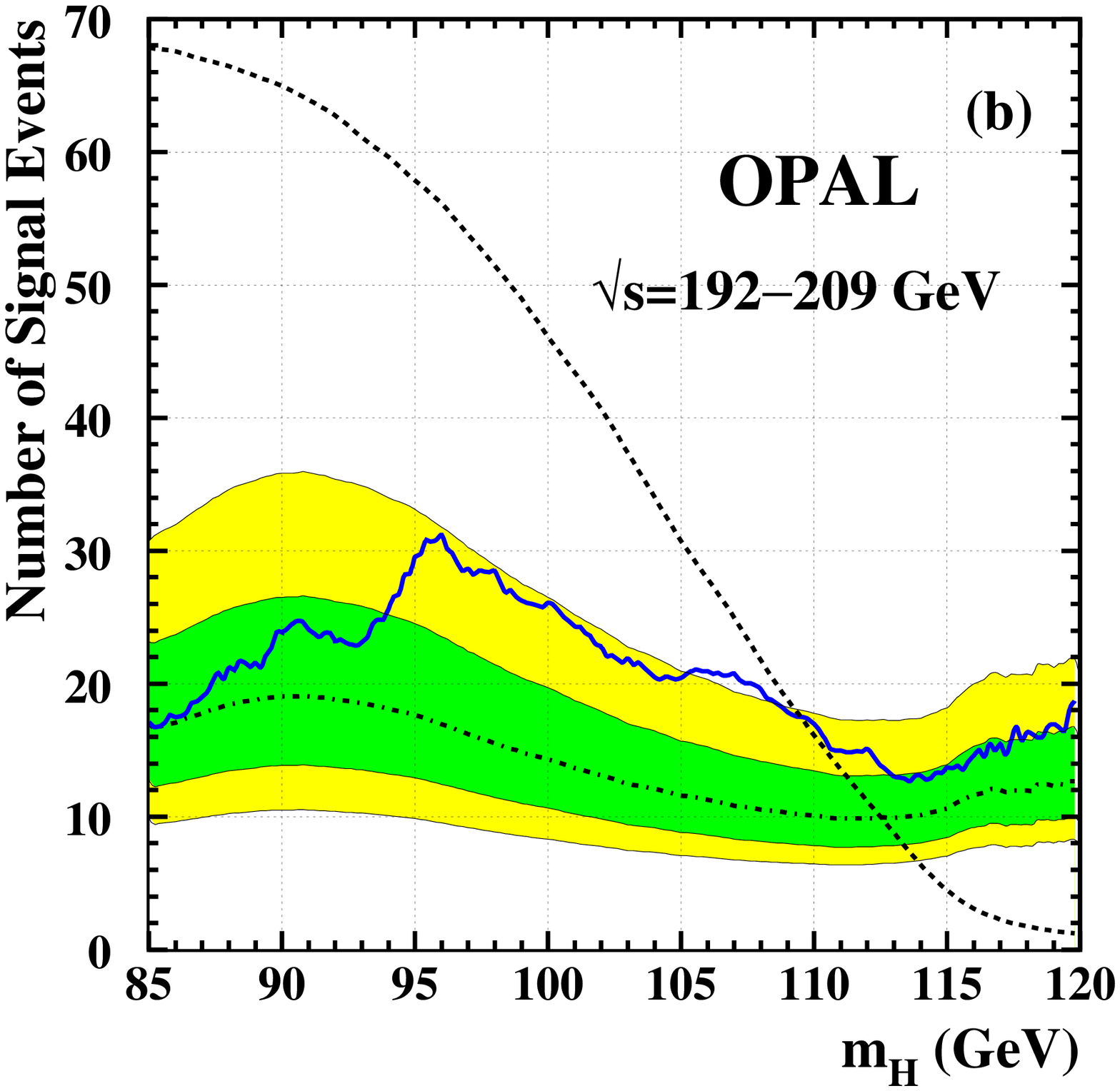}
    \caption{\sl
      (a) The background confidence level $1-$\rm{CL}$_\mathrm{b}$, as
      a function of the Higgs boson test mass. The dotted curve
      represents the median expectation assuming the presence 
      of the Higgs boson with a 115 GeV mass.
      The dark (light) shaded bands indicate the 68\% (95\%) probability
      intervals.
      (b) Upper limits on the signal counts
      at the 95\% confidence level, $n_{95}$, as observed (solid line) 
      and the expected median (dot-dashed line) for expected
      background as a function of the Higgs boson test mass.
      The expected rate of the accepted signal counts 
      for a SM Higgs boson is shown with the dotted line.}
    \label{fig:confidence}
  \end{center}
\end{figure}

The Figure\,\ref{fig:confidence}(a) shows 1$-$CL$_\mathrm{b}$ as a
function of the test mass. An excess can be seen in a range of 95--120
GeV with the lowest value of 0.02 at 107 GeV\@.
The probability to observe such an excess anywhere in the range
of test masses between 95 and 120\,GeV is approximately 10\%, estimated
from the size of the range and the reconstructed mass resolution.
The expected curve in the presence of 115 GeV Higgs is also seen in
the figure. 

The signal rate limit $n_\mathrm{95}$ observed in data is shown as a
function of $m_\PHz$ in Figure\,\ref{fig:confidence}(b) together with
its median expectation in  background-only hypothesis.
Figure\,\ref{fig:confidence}\,(b) also shows the expected accepted
signal rate.   
A lower mass bound of 109.7\,GeV was obtained, and
the expected limit was 112.5\,GeV\@.

\section{Summary}
\label{sec:summary}

A search for the Standard Model Higgs boson has been performed 
with the OPAL detector at LEP based on the full data sample
collected at $\sqrt{s}\approx$192--209\,GeV in 1999 and 2000.
The largest deviation with respect to the expected SM background
in the background confidence level, 1$-$CL$_\mathrm{b}$, 
is 0.02 at 107 GeV\@. The observed excess is less
significant than that expected for a SM Higgs boson with a 107 GeV mass.
A lower bound of 109.7\,GeV on the mass of the SM Higgs boson is obtained
at the 95\% confidence level while the median expectation for
the background-only hypothesis is 112.5 GeV\@.
For a Higgs boson with a mass of 115\,GeV,
1$-$CL$_\mathrm{b}$ is 0.2 while
CL$_\mathrm{s+b}$ is 0.4, indicating that the data slightly
favour the hypothesis that a signal is present, but also that the
data are consistent with the background hypothesis.


\end{document}